\documentstyle [12pt,epsf]{article}

\input{epsf}
\begin{document}
\begin{titlepage}
\begin{center}

\vspace{-0.7in}

{\large \bf Finite Size Effects in \\
the Anisotropic
$\frac{\lambda}{4!}(\varphi^{4}_{1}+\varphi^{4}_{2})_{d}$ Model}\\
\vspace{.3in}{\large\em C.~D.~Fosco}\\
Centro At\'omico Bariloche, 8400 Bariloche, Argentina
\\
and
\\
\vspace{.15in}{\large\em N.~F.~Svaiter}\\
Centro Brasileiro de Pesquisas F\'\i sicas-CBPF\\ 
Rua Dr. Xavier Sigaud 150,
Rio de Janeiro, RJ, 22290-180, Brazil
\vspace{-0.7in}

\subsection*{\\Abstract}
\end{center}
We consider the $\frac{\lambda}{4!}(\varphi^{4}_{1}+\varphi^{4}_{2})$
model on a d-dimensional Euclidean space, where all but
one of the coordinates are unbounded. Translation invariance along the bounded
coordinate, $z$, which lies in the interval $[0,L]$, is broken because 
of the boundary conditions (BC's) chosen for the hyperplanes $z=0$ and
$z=L$.
Two different possibilities for these BC's boundary conditions 
are considered: $DD$ and $NN$, where D denotes Dirichlet and N Newmann,
respectively. The renormalization procedure  up to one-loop order is applied,
obtaining two main results. The first is the fact that the renormalization
program requires the introduction of counterterms which are surface
interactions. The second one is that the tadpole graphs for $DD$ and $NN$
have the same $z$ dependent part in modulus but with opposite signs.
We investigate the relevance of this fact to the elimination of surface
divergences.

\end{titlepage}
\newpage\baselineskip .37in
\section{Introduction}

In this paper we consider an interacting quantum field theory model in 
the presence of boundaries. We shall assume that the system is finite
along one dimension $z \in [0,L]$, and infinitely extended along
the remaining $(d-1)$ directions.
 
The presence of geometric restrictions on the domain of one of the 
coordinates of the system, demands the introduction of classical 
boundary conditions, to be satisfied by the fields on the two hypersurfaces 
at $z=0$ and $z=L$. 
If we restrict ourselves to a real scalar field, Hermiticity of
the Hamiltonian leads us to five different (inequivalent) choices for 
the BC's, namely: $DD$, $NN$, $DN$, periodic and anti-periodic. 
The last two choices are usual in the finite-temperature literature, 
and shall not be dealt with here, since they don't break translation 
invariance, which is the phenomenon we are concerned with.

Physical systems will be, in general, finite along several
directions. For the sake of simplicity we will consider a d-dimensional
layered geometry. Although the highly idealized case of planar boundaries
misses a whole series of features that are present in the general, 
curved boundary case, for more general shapes the multiple reflection 
method can be used to find the correlation functions of the 
model~\cite{balian}. 

Most of the papers in the literature deal with periodic or anti-periodic 
boundary conditions, where translation symmetry is maintained, and
surface effects avoided. Moreover, in quantum systems where translation
symmetry is broken, the renormalization procedure is more involved than for
translation invariant systems, either bounded or unbounded.

The diagrammatic expansion and the renormalization program for an unbounded
system is conveniently performed in momentum space. 
On the other hand, when $DD$ or $NN$ BC's are implemented, one may still
work with Green's functions at fixed $(d-1)$ dimensional momenta, since there
is translation symmetry along those dimensions.
As discussed by many authors, associated with the breaking of 
translational invariance, a new feature emerges: the existence of one 
particle reducible primitively divergent diagrams. For example, the bare
two-point function $G_{0}(x,x')$ of the scalar model with zero, one or two 
points on the surfaces have different renormalization constants,
respectively. 

In this paper we will consider a scalar theory subject to two different 
classical BC's: $DD$ and $NN$. Besides the lack of translational invariance,
we shall face the problem of surface divergences. One way to avoid them is 
to smooth out the plates surface. But in this case an ambiguity appears,
since loop-graphs will depend on an ad-hoc model assumption, namely, the
particular features of the smooth walls. Consequently, we prefer to maintain
the hard walls assumption. In the context of the Casimir energy of
minimally coupled scalar fields, many authors used soft, hard and semi-hard
BC's~\cite{hard}. Different questions sometimes require more 
complicated BC's, like the quantum mechanical treatment of the boundary 
conditions presented by Ford and Svaiter~\cite{flu}, a device implemented to
solve a long standing paradox concerning the renormalized energy of 
minimally and conformally coupled scalar fields.

Besides the above mentioned effects due to the existence of surfaces and the
breaking of translation symmetry, we do also have, of course, finite size
effects, which are of a different nature. Various investigations have been 
made on this subject, mostly from the quantum statistical mechanics point 
of view, and we present a short review here. 
Pathria and co-workers studied the Bose-Einstein condensation of an 
ideal relativistic Bose-gas confined to a rectangular box of sizes 
$L_{1},L_{2}$ and $L_{3}$ with periodic boundary conditions on all the 
walls~\cite{bose}. 
A systematic study of finite size systems and phase transitions was developed 
by Brezin and Zinn-Justin \cite{brez}. These authors studied
the $O(N)$ model in two different geometries: the periodic cube and the 
cylinder along one dimension (the time) and finite and periodic in the $(d-1)$ 
remaining dimensions. Nemirovsky and Freed considered the same model
but in a `complementary' situation regarding the boundary conditions, namely, 
$(d-1)$ dimensions are unbounded, and periodicity along the only finite 
dimension, and \cite{fred}. 
Afterwards, Singh and Pathria studied the $O(N)_{d}$ model confined to 
geometries with periodic boundary conditions in all directions \cite{sing}. 
The same model in the presence of one mirror localized at $z=0$ was 
also studied by many authors \cite{mir}. 

Another ingredient, important in finite size systems at criticality, is the
concept of finite size scaling. Let us consider a finite system of linear size 
$L$ and suppose a thermodynamic quantity $P_{L}(t)$ (where $t$ is the reduced 
temperature) becomes singular as $t\rightarrow 0$. Defining 
$\frac{P_{L}(t)}{P(t)}=g(L,t)$, where $P(t)$ is the bulk 
value of $P_{L}(t)$, the statement of finite size scaling is that 
$g(L,t)=f(L/\xi(t))$, where $\xi(t)$ is the correlation length. 
In other words, finite size scaling predicts that, for large $L$, the 
dependence on $L$ of the singular contributions to thermodynamics functions 
scales with the correlation length, and is described by universal scaling 
functions. It may be pointed out that finite size systems must be classified 
in two distinct groups with respect to finite size scaling,
depending on whether the B.C.'s break translation invariance or not.
For finite size systems where translation invariance is maintained 
(for example, a periodic cube, or a cylinder infinite along one dimension, 
and finite and periodic in the $(d-1)$ other dimensions), finite size 
scaling is easily understood.

The renormalization group equations are insensitive to finite 
size effects (since the renormalization is related to short distance 
singularities) and must, accordingly, be maintained in such finite size 
geometries. However, the solutions to those equations must be different from 
those for the unconstrained systems, because correlation functions can depend 
on the additional dimensional quantities (the lengths of the compactified 
dimensions) and finite size scaling is present. 

For the cases of $DD$, $NN$ or $DN$ B.C.'s, the situation is quite different,
since it is much more difficult to decide if a given interaction is relevant, 
irrelevant or marginal, the reason being that the propagator of the 
critical theory satisfies B.C.'s  which can interfere with the power 
counting. 

Finite size effects have also been extensively studied in the quantum field theory context during the last twenty years. In flat spacetime with one compactified dimension, the mass can depend upon the periodicity length~\cite{in}. This phenomenon is of particular interest in theories with broken symmetry, as it allows topological effects to play a role in the restoration of symmetry. 
An equivalent mechanism is at work when we assume that the fields are in 
thermal equilibrium with a reservoir at temperature $\beta^{-1}$ 
\cite{linde}. Finite size effects in quantum field theory with periodic 
boundary conditions in the spatial section at finite temperature was 
analyzed by many authors. See, for example, \cite{ori}, and
references therein. 

In the context of non-Abelian gauge field theories at zero temperature, 
cavity QCD was studied by many authors~\cite{mit}. Hanson and Jaffe 
and also Hanson, Jonhson and Peterson dealt with quantum fields in  
bounded domains with broken translation invariance.  
Regarding Abelian gauge theories, quantum electrodynamics in 
the presence of conducting plates has also been the focus of research~\cite{bor}. 
Based on the fact that the B.C.'s  for the electron field could lead to 
additional contributions to the Casimir force,  
Bordag et al. and also Robaschik et al. adopted the following model. A 
photon field obeys classical B.C.'s on perfectly conducting 
plates, while the B.C's for the fermion field are free. These authors 
assumed that the electromagnetic field also exists in the region outside 
the plates (two simple connected domains). Chodos and Thorn investigated the 
self-energy of fermions used different B.C's (the slab-bag),  where the 
fermionic field is confined between two parallel plates and the photon field is 
unconfined. In flat spacetime, for systems where some dimensions are 
compactified but translational invariance is maintained, Toms~\cite{pro1} 
and also Birrel and Ford~\cite{pro2} have shown  that all the counterterms 
are independent on the compactified spatial size. A more general discussion has 
been given by Banach~\cite{bach}. This author proved that a topological 
identification (periodic or anti-periodic B.C.'s) does not introduce new 
counterterms into the theory. As stressed by many authors,  were this not 
the case there would be a catastrophe in the renormalizability of the model.

For translation invariant systems, because of  Poincar\'e invariance, one 
should expect that overlapping divergences will  not obstruct the 
implementation of the renormalization program \cite{ward}. In systems where Poincar\'e
invariance does not hold, these proofs do not apply, and one must show 
that it is still possible to implement such program. 
A technical difficulty is also met here, since the presence of
geometric restrictions makes Feynman diagrams harder to compute than 
is ordinary quantum field theory in unbounded systems. 
An crucial work on this subject has been presented by  Symanzik~\cite{zi}. 
Of particular importance are also  the papers by  Nemirovsky and 
Freed, and Krech and Dietrich~\cite{ne}.

In this work we shall consider an anisotropic scalar model, in a d-dimensional 
Euclidean space, where the first $(d-1)$ coordinates are unbounded and the 
last one lies in the interval $[0,L]$. 
We analyze two different translation symmetry breaking B.C.'s:
$DD$ and $NN$ on the plates. 
We first present a rederivation of  the fact that to renormalize the theory 
one has to introduce counterterms as surface interactions. We also show   
that the tadpole graphs for $DD$ and for $NN$ B.C.'s have the same modulus
for their $z$-dependent parts, but their signs are opposite. We 
study the possible use of this property to get rid of the surface 
divergences.

The organization of the paper is as follows: In section II we present  the 
general formalism. In section III we discuss the slab configurations, dealing
with the two-point and four-point functions, both for $DD$ and $NN$ b.c.
In section IV we analyze the divergences of the translational invariant 
part of the tadpoles. Section V deals with the  analysis of the ultraviolet 
and infrared divergences of the $z$-dependent part of the tadpoles.
Finally, section VI contains our conclusions. 
Throughout this paper we use $\hbar=c=1$.

\section{ General formalism and the scalar anisotropic model}\

Let us consider $Z[J]$, the generating functional of complete Green's  
functions for a scalar field in a $d$-dimensional Euclidean space
\begin{equation}
Z[J]= \int{\cal D}\varphi\ e^{-S[\varphi]+ \int J(x)\varphi(x)}
\label{outin}
\end{equation}
where
\begin{equation}
S[\varphi]=\int d^{d}x[{\cal L}(\varphi(x),\partial\varphi(x))]\,,
\label{action}
\end{equation}
${\cal D}\varphi$ is the appropriate measure, and $ S[\varphi]$ is the 
classical action associated with the scalar field. The quantity $Z[J]$ 
can be regarded as a functional integral representation for the 
imaginary time evolution operator 
$\left<\varphi_2\right|U(t_2,t_1)\left|\varphi_1\right>$
with the boundary conditions: $\varphi(t_1,x)=\varphi_1(\vec{x})$
and $\varphi(t_2,x)=\varphi_2(\vec{x})$.
The quantity $Z[J]$ gives the transition amplitude from the initial
state $\left|\varphi_1\right>$ to the final state $\left|\varphi_2\right>$
in the presence of some scalar source $J(x)$, of compact support.
Regarding the  Lagrangian density ${\cal L}$, we shall assume it to be 
\begin{equation}
{\cal L}(\varphi) =\frac{1}{2}(\partial\varphi)^{2}+
\frac{1}{2}m^{2}\varphi^{2}+\frac{1}{4!}\lambda\varphi^{4} \;.
\label{lag}
\end{equation}
The n-point correlation functions are given by the expectation 
value with respect to the weight $e^{-S(\varphi)}$ defined as
\begin{eqnarray}
G^{(n)}(x_{1},x_{2},..,x_{n})=<\varphi(x_{1})...\varphi(x_{n})>&=&
\frac{1}{Z(J)}\frac{\delta^{n}Z(J)}
{\delta J(x_{1})\delta J(x_{2})..\delta J(x_{n})}|_{J=0} \nonumber \\
& & \nonumber \\
&=&\int{\cal D}\varphi\varphi(x_{1})...\varphi(x_{n})
 e^{-S[\varphi]}\,.
\label{nn}
\end{eqnarray}
As usual,  $W(J)$, the generating functional for connected correlation 
functions of the elementary fields shall be given by $W(J)=ln Z(J)$. 
Thus
\begin{equation}
G_{c}^{(n)}(x_{1},x_{2},..,x_{n})=
\frac{\delta^{n}W(J)}
{\delta J(x_{1})\delta J(x_{2})..\delta J(x_{n})}|_{J=0}=
<\varphi(x_{1})...\varphi(x_{n})>_{c}.
\label{n-con}
\end{equation}
Finally, $\Gamma(\varphi_{0})$, the generating functional of  connected one-particle irreducible correlation functions is introduced by performing a Legendre 
transformation on $W[J]$,
\begin{equation}
\Gamma(\varphi_{0})= -W(J)+\int d^{d}x\,\varphi(x)J(x)
\label{gen-con}
\end{equation}
and 
\begin{equation}
\Gamma^{(n)}(x_{1},x_{2},..,x_{n})=
\frac{\delta^{n}\Gamma(\varphi_{0})}
{\delta \varphi_{0}(x_{1})\delta \varphi_{0}(x_{2})..
\delta \varphi_{0}(x_{n})}|_{\varphi_{0}=0}
\label{1PI}
\end{equation}
where
\begin{equation}
\varphi_{0}(x)=\frac{\delta W}{\delta J(x)}.
\label{ether}
\end{equation}
If $\lambda=0$ the partition function 
$Z(J)$ can be calculated exactly i.e.
\begin{equation}
Z_{0}(J)=
exp\left(\frac{1}{2} \int d^{d}x d^{d}y\, J(x)D(x-y,m^{2})J(y)\right),
\label{exact}
\end{equation}
where
\begin{equation}
(-\Delta_{x}+m^{2})D(x-y,m^{2})=\delta^{d}(x-y).
\end{equation}
For $\lambda\neq 0$ 
it is not possible to find exactly $Z(J)$ and perturbation 
theory is mandatory. This expansion stems from the formal identity: 
\begin{equation}
Z(J)\,=\,
exp\left[-\frac{\lambda}{4!}
\int d^{d}x\left(\frac{\delta}{\delta J(x)}\right)^{4}\right]
Z_{0}(J).
\label{pert}
\end{equation}

From now on, we shall consider a generalization of the previous case,
namely, the anisotropic Landau-Ginzburg model for a $N=2$ component 
order parameter with a Lagrange density ${\cal L}=
{\cal L}_{0}+{\cal L}_{int}$,
where
\begin{equation}
{\cal L}_{0}(\varphi_{1},\varphi_{2}) =\frac{1}{2}(\partial\varphi_{1})^{2}+
\frac{1}{2}(\partial\varphi_{2})^{2}+\frac{1}{2}m^{2}\varphi_{1}^{2}+
\frac{1}{2}m^{2}\varphi_{2}^{2}
\label{L0}
\end{equation}
and
\begin{equation}
{\cal L}_{int}=\frac{\lambda}{4!}(\varphi_{1}^{4}
+\varphi_{2}^{4}).
\label{LI}
\end{equation}
To generate the n-point functions we have to introduce two 
scalar sources $J_{1}(x)$ and $J_{2}(x)$ coupled linearly with the 
fields $\varphi_{1}(x)$ and $\varphi_{2}(x)$ respectively. Integrating
out the fields, we obtain $Z(J_{1})$ and  $Z(J_{2})$ and 
the total partition function of the model factorizes: 
$Z(J_{1},J_{2})=Z(J_{1})Z(J_{2})$, where
\begin{equation}
Z(J_{1,2})=\frac{1}{N_{1,2}}
exp\left(-\frac{\lambda}{4!}
\int d^{d}x\left(\frac{\delta}{\delta J_{1,2}(x)}\right)^{4}\right)
Z_{0}(J_{1,2}).
\label{1}
\end{equation}
and 
\begin{equation}
Z_{0}(J_{1,2})=
exp\left(\frac{1}{2} \int d^{d}x d^{d}y\, 
J_{1,2}(x)G_{1,2}^{(2)}(x-y,m^{2})J_{1,2}(y)\right),\;.
\end{equation}
In the above equation, $G_{1}^{(2)}(x-y,m^{2})$
and $G_{2}^{(2)}(x-y,m^{2})$ are the free propagators, 
solutions of the inhomogeneous equations $(i=1,2)$.
\begin{equation}
(-\Delta_{x}+m^{2})G_{i}^{(2)}(x-y,m^{2})=\delta^{d}(x-y).
\end{equation} 
The partition function applies to arbitrary geometries, and 
classical B.C.'s must be implemented on the Green's functions. 
As discussed before, we will assume that the system is confined between 
two parallel plates localized at $z=0$ and $z=L$, and use the 
Cartesian coordinates $x^{\mu}=(\vec{r},z)$ where $\vec{r}$ is a $(d-1)$ 
dimensional vector perpendicular to the $z$ direction. A question that 
arises in such a model is related to the renormalization conditions. 
It is well known that for fields interacting with a thermal bath
defined in manifolds where the spacelike 
sections are non-compact the mass and coupling constant counterterms are 
temperature independent. Using dimensional regularization~\cite{dim}
it was proved that 
for fields defined on manifolds where the spacelike sections are non-compact, 
or compact in at least one dimension, but with the other dimensions 
noncompactified, the mass and coupling constant counterterms are size and
temperature independent at the two-loop level. 
In a perturbative scheme, the renormalized theory 
is fixed by the renormalization conditions for the superficially divergent 
vertex functions (the one particle irreducible parts of the 
connected Green's functions). In other words, in the conventional 
renormalizable (translational invariant) theory the ultraviolet 
divergences can be absorbed 
by counterterms related to the field, mass and coupling constant. 
A question of fundamental importance is how the renormalization 
program can be implemented in systems where translational 
invariance is broken. The purpose of the next section 
is to analyze this question for the case of  the anisotropic model at the 
one-loop approximation.

\section{Finite size effects and classical boundary conditions}\

For the cubic anisotropic model, we define the  boundary 
conditions over the plates for the fields $\varphi_{1}(x)$ and 
$\varphi_{2}(x)$.
For the $\varphi_{1}(x)$ field we assume Dirichlet-Dirichlet boundary
conditions i.e:
\begin{equation}
\varphi_{1}(\vec{r},z)|_{z=0}=\varphi_{1}(\vec{r},z)|_{z=L}=0,
\label{dir}
\end{equation}
and for the $\varphi_{2}(x)$ we will assume Newmann-Newmann boundary
conditions, i.e.:
\begin{equation}
\frac{\partial}{\partial{z}}\varphi_{2}(\vec{r},z)|_{z=0}
=\frac{\partial}{\partial{z}}\varphi_{2}(\vec{r},z)|_{z=L}=0.
\label{new}
\end{equation}

It is well known that ${^{4}He}$ films close to the $\lambda$ transition 
satisfies $DD$ b.c. Another well known example of such kind of boundary conditions 
is the electromagnetic field. It was shown that for an electromagnetic 
field confined in 
a perfectly conducting cavity, it is possible to treat 
the electric and magnetic modes separately, where each one
satisfies Dirichlet and Newmann B.C.'s, respectively~\cite{Greiner}. 
Going back to our discussion, since 
the translational invariance is not preserved, let us use 
a Fourier expansion of the fields in the following form:
\begin{equation}
\varphi(\vec{r},z)=\frac{1}{(2\pi)^{\frac{d-1}{2}}}
\int d^{d-1}p\sum_{n}\phi_{n}(\vec{p})e^{i\vec{p}.\vec{r}}
u_{n}(z).
\label{field}
\end{equation}
where the $u_{n}(z)$ are the normalized eigenfunctions 
of the operator $-\frac{d^2}{dz^{2}}$ satisfying the  
completeness and orthonormality relations, i.e.,
\begin{equation}
\sum_{n}u_{n}(z)u_{n}^{*}(z')= \delta(z-z'),
\label{comp}
\end{equation}
\begin{equation}
\int_{0}^{L} dz\, u_{n}(z)u_{n'}^{*}(z)= \delta_{n,n'},
\label{orto}
\end{equation}
and finally
\begin{equation}
-\frac{d}{dz^{2}}u_{n}(z)=k^{2}_{n}u_{n}(z),
\label{ult}
\end{equation}
where $k_{n}=\frac{n\pi}{L}$, $n=1,2..$ for $DD$ b.c and 
$n=0,1,2..$ for $NN$ b.c.

It should be noted that this kind of expansion in an orthonormal set
corresponding to the eigenfunctions of the Hermitian operator 
$-\frac{d^2}{dz^{2}}$ defined on a finite interval could be quite
straightforwardly generalized to different anisotropic models. For
example, we might consider a scalar field defined on Euclidean
space with all the coordinates unbounded, but with a mass having
an anisotropy along the $z$ coordinate. Namely, the Lagrangian density
could be
\begin{equation}
{\cal L}\;=\; \frac{1}{2} (\partial \varphi)^2 + \frac{1}{2} (m^2 +
\mu^2(z) ) \varphi^2 + \frac{\lambda}{4!} \varphi^4 \;, 
\end{equation} 
which is, of course, non invariant under translations in $z$,
except for the trivial case of a constant $\mu (z)$.
If now we assume the $u_n(z)$'s to denote the normalized eigenfunctions of the 
Hermitian operator
\begin{equation}
h \;=\; -\frac{d^2}{dz^2} + \mu (z)\;,
\end{equation}
with  
\begin{equation}
h \, u_n (z) \;=\; \lambda_n^2 u_n (z) \,,
\end{equation}
the expansion (\ref{field}) still holds. We assume $\mu$ to be a non negative
function, so that $h$ is definite positive.

We see that the case of Dirichlet B.C.'s could be obtained starting from an 
unbounded $z$ coordinate, and using the anisotropic mass 
$\mu(z) = \mu (\frac{z}{L})^n$, with $n \to \infty$, and $\mu$ a (positive) 
constant. For a study of Dirac fermions with a space dependent mass see
for example ref. \cite{ana}.

Coming back to the case of $DD$ and $NN$ boundary conditions, the 
eigenfunctions are, respectively,
\begin{equation}
u_{n}(z)=\sqrt{\frac{2}{L}}\sin(\frac{n\pi z}{L})\,\,\, n=1,2..
\label{dirmode}
\end{equation}
and 
\begin{equation}
u_{n}(z)=\sqrt{\frac{2}{L}}\cos(\frac{n\pi z}{L})\,\,\, n=1,2,\cdots \;.
\label{newmode}
\end{equation}
For $NN$ B.C.'s,  we also have the zero mode 
$u_{0}(z)=\frac{1}{\sqrt{L}}$.
The free propagator can be expressed in the following form:
\begin{equation}
G_{0}^{(2)}(\vec{r},z,z')=\frac{1}{(2\pi)^{d-1}}\int d^{d-1}p\sum_{n}
e^{i\vec{p}.\vec{r}}u_{n}(z)u_{n}^{*}(z')\,G_{0,n}(\vec{p}),
\label{ge}
\end{equation}
where is not difficult to show that $G_{0,n}(\vec{p})$ is given by
\begin{equation}
G_{0,n}(\vec{p})=(\vec{p}^{\,2}+k_{n}^{2}+m^{2})^{-1}.
\label{geex}
\end{equation}
For the anisotropic mass case, we would have instead:
\begin{equation}
G_{0,n}(\vec{p})=(\vec{p}^{\,2}+\lambda_{n}^{2}+m^{2})^{-1}.
\end{equation}

As we discussed before, for translational invariant systems we 
have $G_{0}^{(2)}(x,x')=G_{0}^{(2)}(x-x')$ and from coordinate space Feynman 
rules we can go to momentum space representation, which is 
the more convenient framework to analyze the divergences of the theory. 
Translation invariance is reflected in momentum conservation conditions. Since 
our system possesses translation invariance along the direction parallel
to the plates, the parallel momentum, $\vec{p}$, is conserved.
In this case a convenient representation is a mixed $(\vec{p},z)$ 
space. The Feynman rules for different boundary conditions was 
derived in many references and we will not repeat the rules here.
For a careful study of Feynman rules in such 
systems see for example Ref. \cite{Dil}.
Let us study the one-loop correction to the bare two-point 
function $G_{0}^{(2)}(x,x')$, both for the 
$DD$ and $NN$ cases. Using the Feynman rules (see fig.(1)) we have: 
\begin{equation}
G_{0}^{(2)}(\lambda,\vec{r}_{1},z_{1},\vec{r}_{3},z_{3}) =\frac{\lambda}{2}\int d^{d-1}r_{2}\int_{0}^{L}dz_{2}G_{0}^{(2)}(\vec{r}_{1},z_{1};\vec{r}_{2},z_{2})
G_{0}^{(2)}(\vec{r}_{2},z_{2};\vec{r}_{2},z_{2})
G_{0}^{(2)}(\vec{r}_{2},z_{2};\vec{r}_{3},z_{3}).
\label{M}
\end{equation}
or 
\begin{equation}
G_{0}^{(2)}(\lambda,\vec{r}_{1},z_{1},\vec{r}_{3},z_{3}) =\frac{\lambda}{2}\int d^{d-1}r_{2}\int_{0}^{L}dz_{2}G_{0}^{(2)}(\vec{r}_{1}-\vec{r}_{2};z_{1},z_{2})
G_{0}^{(2)}(\vec{0},z_{2})G_{0}^{(2)}(\vec{r}_{2}-\vec{r}_{3};z_{2},z_{3}).
\label{MF}
\end{equation}
\begin{figure}[ht]
  
\centerline{\epsfysize=1.0in\epsffile{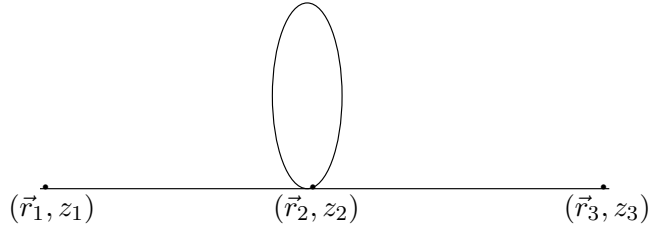}} 
\caption[region]
{The two point function for $\varphi_1(x)$}

\begin{picture}(10,10)
\put(130,50){{\small $(\vec{r}_1,z_1)$}}

\put(230,50){{\small $(\vec{r}_2,z_2)$}}

\put(340,50){{\small $(\vec{r}_3,z_3)$}}

\end{picture}

\end{figure}

Each of these expressions can, for the case 
of $DD$ B.C.'s, be expanded as
\begin{equation}
G_{0}^{(2)}(\vec{r}_{1}-\vec{r}_{2},z_{1},z_{2})=\frac{2}{L}\frac{1}{(2\pi)^{d-1}}
\sum_{n=1}^{\infty}\sin(\frac{n\pi z_{1}}{L})\sin(\frac{n\pi z_{2}}{L})
\int d^{d-1}p\frac{e^{i\vec{p}.(\vec{r}_{1}-\vec{r}_{2})}}
{(\vec{p}^{\,2}+(\frac{n\pi}{L})^{2}+m^{2})}.
\label{g1}
\end{equation}
\begin{equation}
G_{0}^{(2)}(\vec{r}_{2}-\vec{r}_{3},z_{2},z_{3})=\frac{2}{L}\frac{1}{(2\pi)^{d-1}}
\sum_{n'=1}^{\infty}\sin(\frac{n\pi z_{2}}{L})\sin(\frac{n\pi z_{3}}{L})
\int d^{d-1}p\frac{e^{i\vec{p}.(\vec{r}_{2}-\vec{r}_{3})}}
{(\vec{p}^{\,2}+(\frac{n'\pi}{L})^{2}+m^{2})},
\label{g2}
\end{equation}
and finally
\begin{equation}
G_{0}^{(2)}(\vec{0},z_{2})=\frac{2}{L}\frac{1}{(2\pi)^{d-1}}
\sum_{n''=1}^{\infty}\sin^{2}\frac{n''\pi z_{2}}{L}
\int d^{d-1}p\frac{1}
{(\vec{p}^{\,2}+(\frac{n''\pi}{L})^{2}+m^{2})}\;.
\label{g0}
\end{equation}
Although the functions $G_{0}^{(2)}(\vec{r}_{1}-\vec{r}_{2},z_{1},z_{2})$ and 
$G_{0}^{(2)}(\vec{r}_{2}-\vec{r}_{3},z_{2},z_{3})$ are singular at $\vec{r}_{1}=\vec{r}_{2}$, $z_{1}=z_{2}$ and 
$\vec{r}_{2}=\vec{r}_{3}$, $z_{2}=z_{3}$, the singularities are 
integrable (for points outside the plates), consequently only the 
tadpole is divergent and needs a 
regularization and renormalization procedure.
A straightforward calculation yields the order $\lambda$ correction to 
the bare two-point function in the one-loop approximation :
\begin{eqnarray}
G_{0}^{(2)}(\lambda,\vec{r}_{1}-\vec{r}_{3},z_{1},z_{3})& & =
\frac{2}{L^{2}}\frac{1}{(2\pi)^{d-1}}\int_{0}^{L}dz_{2}
\sum_{n,n'=1}^{\infty}
\sin(\frac{n\pi z_{1}}{L})
\sin(\frac{n\pi z_{2}}{L})
\sin(\frac{n'\pi z_{2}}{L})
\sin(\frac{n'\pi z_{3}}{L}) \nonumber \\
& & \int d^{d-1}p \frac{e^{i\vec{p}(\vec{r}_{1}-\vec{r}_{3})}}
{ (\vec{p}^{\,2}+(\frac{n\pi}{L})^{2}+m^{2})  (\vec{p}^{\,2}+(\frac{n'\pi}{L})^{2}+m^{2})} 
 T_{DD}(L,m,d,z_{2}) 
\label{grande}
\end{eqnarray}
where, since we will use dimensional regularization, we introduce a 
dimensional parameter $\mu$, and define $g=\lambda\mu^{4-d}$. 
The expression for the tadpole $T_{DD}(L,m,d,z)$ is then:
\begin{equation}
T_{DD}(L,m,d,z)=\frac{2g}{L}\frac{1}{(2\pi)^{d-1}}\sum_{n=1}^{\infty}
\sin^{2}(\frac{n\pi z}{L})
\int d^{d-1}p\frac{1}
{(\vec{p}^{\,2}+(\frac{n\pi}{L})^{2}+m^{2})}.
\label{TadDD}
\end{equation} 
The tadpole graph in the case of N-N B.C.'s can be also easily 
found, and it is given by
\begin{eqnarray}
T_{NN}(L,m,d,z)& & =\frac{g}{L}\frac{1}{(2\pi)^{d-1}}\int d^{d-1}k\frac{1}
{(\vec{k}^{2}+m^{2})}  \nonumber \\
& & +\frac{2g}{L}\frac{1}{(2\pi)^{d-1}}\sum_{n=1}^{\infty}
\cos^{2}(\frac{n\pi z}{L}) 
\int d^{d-1}p\frac{1}
{(\vec{p}^{\,2}+(\frac{n\pi}{L})^{2}+m^{2})}.
\label{TdaNN}
\end{eqnarray}
Note that both $T_{DD}(L,m,d,z)$ and $T_{NN}(L,m,d,z)$ 
diverge in their continuum momenta integrals and also
in the $n$ summation. In the next section we will analyze the ultraviolet 
behaviour of the bare two-point functions i.e  $ T_{DD}(L,m,d,z) $  and 
$T_{NN}(L,m,d,z)$ .
Before dealing with the renormalization of the one-loop two point function, 
let us, by the sake of completeness, discuss the bare four-point function.
The expression for the bare four-point function is given below, and 
the same analysis of the divergences can be done. In this paper we will not 
implement the renormalization program of the four point-function 
(which follows using the same procedure used in the two-point function). 
Our object of interest is the two-point function, since it is the fundamental 
quantity that measures the vacuum activity. Using the Feynman rules, 
$G_{0}^{(4)}(\lambda,x_{1},x_{2},x_{3},x_{4})$,
the order $\lambda^{2}$ 
correction to the bare four-point function, is given by
\begin{eqnarray}
&& G_{0}^{(4)}(\lambda,\vec{r}_{1},z_{1},\vec{r}_{2},z_{2},\vec{r}_{5},z_{5},
\vec{r}_{6},z_{6}) =
\frac{1}{2}\int d^{d-1}r_{3}\int d^{d-1}r_{4}
\int_{0}^{L}dz_{3} \int_{0}^{L}dz_{4}\;G_{0}^{(2)}(\vec{r}_{1}-\vec{r}_{3},z_{1},z_{3})
\nonumber \\
&& G_{0}^{(2)}(\vec{r}_{2}-\vec{r}_{3},z_{2},z_{3}) 
[G_{0}^{(2)}(\vec{r}_{3}-\vec{r}_{4},z_{3},z_{4})]^2 
G_{0}^{(2)}(\vec{r}_{4}-\vec{r}_{5},z_{4},z_{5})
G_{0}^{(2)}(\vec{r}_{4}-\vec{r}_{6},z_{4},z_{6}).
\label{nova}
\end{eqnarray}
\begin{figure}[ht]
  
\centerline{\epsfysize=1.0in\epsffile{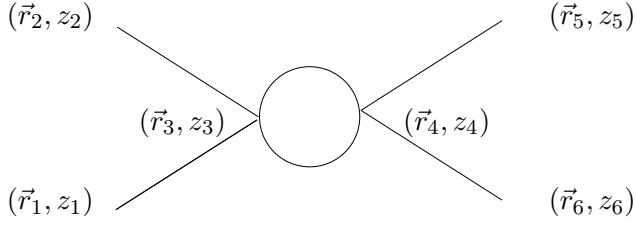}} 
\caption[region]
{The four point function for $ \varphi_1(x) $ }

\begin{picture}(10,10)
\put(135,60){{\small $(\vec{r}_1,z_1)$}}

\put(135,130){{\small $(\vec{r}_2,z_2)$}}

\put(340,60){{\small $(\vec{r}_6,z_6)$}}

\put(340,130){{\small $(\vec{r}_5,z_5)$}}

\put(185,90){{\small $(\vec{r}_3,z_3)$}}
\put(285,90){{\small $(\vec{r}_4,z_4)$}}

\end{picture}

\end{figure}

Again, all $G_{0}$'s are singular at the same points, but the singularities
are integrable, except for $G_{0}^{(2)}(\vec{r}_{3},z_{3},\vec{r}_{4},z_{4})$,
consequently, to renormalize the bare four-point function 
we have the regularize  the 1PI four-point function $\Gamma^{(4)}(\lambda,\vec{r}_{3}-\vec{r}_{4},z_{3},z_{4})=
[G_{0}^{(2)}(\lambda,\vec{r}_{3}-\vec{r}_{4},z_{3},z_{4})]^2$, 
which, for $DD$ B.C.'s, is given by 
\begin{equation}
G_{0}^{(2)}(\vec{r}_{3}-\vec{r}_{4},z_{3},z_{4})=\frac{2g}{L}\frac{1}{(2\pi)^{d-1}}
\sum_{n=1}^{\infty}\sin(\frac{n\pi z_{3}}{L})\sin(\frac{n\pi z_{4}}{L})
\int d^{d-1}p\frac{e^{i\vec{p}.(\vec{r}_{3}-\vec{r}_{4})}}
{(\vec{p}^{\,2}+(\frac{n\pi}{L})^{2}+m^{2})}.
\label{amput}
\end{equation}
A convenient way to express $G_{0}(\vec{r}_{3}-\vec{r}_{4},z_{3},z_{4})$ 
is the following. Let us define
$\vec{\rho}=\vec{r}_{3}-\vec{r}_{4}$, and also $z_{3}-z_{4}=u$,
$z_{3}+z_{4}=v$  then it is possible to write 
\begin{equation}
G_{0}^{(2)}(\vec{r}_{3}-\vec{r}_{4},z_{3},z_{4})=G_{0}^{(2)}(\vec{\rho},u)+
G_{0}^{(2)}(\vec{\rho},v),
\label{uv}
\end{equation}
where
\begin{equation}
G_{0}^{(2)}(\vec{\rho},u)=\frac{g}{L}\frac{1}{(2\pi)^{d-1}}
\sum_{n=1}^{\infty}\cos(\frac{n\pi u}{L})
\int d^{d-1}p\frac{e^{i\vec{p}.\vec{\rho}}}
{(\vec{p}^{\,2}+(\frac{n\pi}{L})^{2}+m^{2})},
\end{equation}
and also 
\begin{equation}
G_{0}^{(2)}(\vec{\rho},v)=-\frac{g}{L}\frac{1}{(2\pi)^{d-1}}
\sum_{n=1}^{\infty}\cos(\frac{n\pi v}{L})
\int d^{d-1}p\frac{e^{i\vec{p}.\vec{\rho}}}
{(\vec{p}^{\,2}+(\frac{n\pi}{L})^{2}+m^{2})}.
\end{equation}
For simplicity let us choose $m=0$ and $\vec{\rho}=0$. Using 
the fact that $d^{d-1}p=p^{d-2}dp\,d\Omega_{d-1}$ and 
$\int \, d\Omega_{d-1}=\frac{2\pi^{\frac{d-1}{2}}}{\Gamma(\frac{d-1}{2})}$,
a straightforward calculation yields
\begin{equation}
G_{0}^{(2)}(\vec{\rho},u,v)|_{\rho=0}=B_{1}(d,L,u)+B_{2}(d,L,v)+
B_{3}(d,L,v)+B_{4}(d,L,u),
\label{bb}
\end{equation}
where
\begin{equation}
B_{1}(d,L,u)=\frac{2g}{L}h_2(d)\int dk\,k^{d-3}\coth Lk \cosh ku.
\label{b1}
\end{equation}
\begin{equation}
B_{2}(d,L,v)=-\frac{2g}{L}h_2(d)\int dk\,k^{d-3}\coth kL \cosh kv.
\label{b2}
\end{equation}
\begin{equation}
B_{3}(d,L,v)=\frac{2g}{L}h_2(d)\int dk\,k^{d-3}\sinh kv.
\label{b3}
\end{equation}
and finally
\begin{equation}
B_{4}(d,L,u)=-\frac{2g}{L}h_2(d)\int dk\,k^{d-3}\sinh ku.
\label{b4}
\end{equation}
It is worth mentioning that the structure of the divergences of 
Eqs.(\ref{b1}-\ref{b4}) are the same as for the tadpoles, as we will see.
In the next section we will analyze the renormalization program 
for the two-point functions in both cases of $DD$ and $NN$ 
boundary conditions.

\section{Analysis of the ultraviolet divergences 
of  $T_{DD}(L,m,d,z)$ and 
$T_{NN}(L,m,d,z)$}

The aim of this section is to analyze the structure of the 
divergences of the bare two-point functions for both cases 
$DD$ and $NN$ boundary conditions.
Let us start from the expression of the vacuum activity for the 
case of $DD$ boundary conditions, i.e.,
\begin{equation}
T_{DD}(L,m,d,z)=\frac{2g}{L}\frac{1}{(2\pi)^{d-1}}\sum_{n=1}^{\infty}
\sin^{2}(\frac{n\pi z}{L})
\int d^{d-1}p\frac{1}
{(\vec{p}^{\,2}+(\frac{n\pi}{L})^{2}+m^{2})}.
\label{TadDD1}
\end{equation} 
Using trigonometric identities and also the relation~\cite{grads}
\begin{equation}
\sum_{n=1}^{\infty}\frac{\cos nx}{n^{2}+a^{2}}=
-\frac{1}{2a^{2}}+\frac{\pi}{2a}\frac{\cosh a(\pi-x)}{\sinh \pi a}
\label{ident}
\end{equation}
which is valid for $0\leq x \leq 2\pi$, it is easy to show that 
the vacuum activity in the case of $DD$ b.c. is given by
\begin{equation}
T_{DD}(L,m,d,z)=
\frac{g}{2L}\frac{1}{(2\pi)^{d-1}}\sum_{n=-\infty}^{\infty}
\int d^{d-1}p\frac{1}
{(\vec{p}^{\,2}+(\frac{n\pi}{L})^{2}+m^{2})}-gf_{2}(d,L,m,z)
\label{f1f2}
\end{equation}
where
\begin{equation}
f_{2}(L,m,d,z)=\frac{1}{2}\frac{1}{(2\pi)^{d-1}}
\int d^{d-1}p\frac{1}
{(\vec{p}^{\,2}+m^{2})^{\frac{1}{2}}}
\frac{\cosh((L-2z)(\vec{p}^{\,2}+m^{2})^{\frac{1}{2}})}
{\sinh (L(\vec{p}^{\,2}+m^{2})^{\frac{1}{2}})}.
\label{f2}
\end{equation}
In an analogous way, it is also possible to calculate the vacuum 
activity for the $NN$ b.c. i.e. 
$T_{NN}(L,m,d,z)$ and we obtain
\begin{equation}
T_{NN}(L,m,d,z)=\frac{g}{2L}\frac{1}{(2\pi)^{d-1}}\sum_{n=-\infty}^{\infty}
\int d^{d-1}p
\frac{1}{(\vec{p}^{\,2}+(\frac{n\pi}{L})^{2}+m^{2})}+gf_{2}(L,m,d,z).
\label{dif}
\end{equation}
Since $T_{DD}(L,m,d,z)$ and $T_{NN}(L,m,d,z)$ have the same functional 
form, both have the same kind of ultraviolet divergences. Let us define
$f_{1}(L,m,d)$ by:
\begin{equation}
f_{1}(L,m,d)= \frac{1}{2L}\frac{1}{(2\pi)^{d-1}}
\sum_{n=-\infty}^{\infty} \int d^{d-1} p 
\frac{1}{(\vec{p}^{\,2}+(\frac{n\pi}{L})^{2}+m^{2})}.
\label{f1}
\end{equation}
The equation above has ultraviolet divergences, but it is (formally)
proportional to the tadpole in finite temperature field theory, 
after the identification: $\beta \equiv 2L$.
To deal with the divergences of the one-loop two-point 
function at finite temperature we have to do frequency sums and 
$(d-1)$ dimensional integrals for the continuum momenta. The most popular 
method to deal with Matsubara sums is to analytic extension away 
from the discrete complex energies down to real axis with the 
replacement of the energy sums by contour integrals \cite{Kapusta}.
We prefer to use dimensional regularization in the continuum~\cite{dim}, 
and afterwards to analytically extend the modified Epstein zeta 
function which appear after the dimensional regularization \cite{lang}.
Since the formalism has already been developed by Malbouisson and 
Svaiter in~\cite{physica}, we will only sketch the procedure here. First we 
have to use a well known result of dimensional regularization, i.e.
\begin{equation}
\int\frac{d^{d}k}{(k^{2}+a^{2})^{s}}=
\frac{\pi^{\frac{d}{2}}}{\Gamma(s)}\Gamma(s-\frac{d}{2})
\frac{1}{a^{2s-d}},
\label{dimensi}
\end{equation}
and let us define the modified Epstein zeta function $\zeta(z,a)$
by:
\begin{equation}
\zeta(z,a)=\sum_{n=-\infty}^{\infty}\frac{1}{(n^{2}+a^{2})^{z}}
\,\,\,\,  a^{2}>0, 
\label{zeta}
\end{equation}
which is analytic for $Re(z)>\frac{1}{2}$. It is possible to 
analytic extend the modified Epstein zeta function where 
the integral representation is valid for $Re(z)<1$, 
\cite{exten}:
\begin{equation}
\sum_{n =-\infty}^{\infty} \bigl( n^2 +a^2 \bigr)^{-z}
= a^{1-2z} \Biggl[ \sqrt{\pi}\, {{\Gamma(z-{1\over 2})} \over
{\Gamma(z)}} 
+ 4 \sin{\pi z} \int_1^{\infty} 
{{(t^2-1)^{-z} dt} \over {{\rm e}^{2\pi a t} -1}} \Biggr] \, . 
\label{sumfor}
\end{equation}
Using Eqs.(\ref{dimensi}) and (\ref{sumfor}) in Eq.(\ref{f1}),
we get a polar part (size independent)
plus a size dependent analytic correction. 
It is clear that the mass counterterm generated by $f_{1}(L,m,d)$
is size independent, as the 
finite temperature field theory has no temperature dependent counterterm. 
The first interesting result of the paper is given by 
Eqs.(\ref{f1f2}) and 
and (\ref{dif}). The tadpole graphs expressed by $T_{DD}(L,m,d,z)$ 
and $T_{NN}(L,m,d,z)$ have the same $z$ dependent part in 
modulus but with opposite signs. From the above discussion 
it is possible to understand the finiteness of the renormalized
stress-tensor of an electromagnetic 
field near a flat prefectly conducting plate. Although the expectation 
value of the squared electric and magnetic field are divergent, a delicate 
cancellation makes the renormalized stress-tensor finite. 
As the size-dependent parts of $T_{DD}(L,m,d,z)$ and $T_{NN}(L,m,d,z)$ have the same functional form and opposite signs, and recalling that it is possible to treat the electric and magnetic modes
separately (where one obeys $DD$ and the other $NN$ b.c., we
automatically obtain a finite result for the vacuum expectation 
value of the stress-tensor of the electromagnetic field. 
It is important to stress that when the conducting boundary is curved,
the energy density diverges on the boundary \cite{Candelas}.

To shall deal with the renormalization program in the one-loop 
approximation in the next section, also discussing, for the sake of completeness, the issue of IR divergencies in different numbers of dimensions.

\section{Analysis of the ultraviolet and infrared divergences of the
$z$-dependent part of the tadpoles}

We will again use the fact that $d^{d-1}p=p^{d-2}dp\,d\Omega_{d-1}$ and 
$\int \, d\Omega_{d-1}=\frac{2\pi^{\frac{d-1}{2}}}{\Gamma(\frac{d-1}{2})}$.
It should be 
noted that, had we chosen $m^{2}=0$, the ultraviolet divergences would have 
kept the same polar structure. Consequently, for simplicity let us choose 
again $m=0$, and for reasons that will become evident latter, we 
first assume $d>3$. The special case $d=3$ is discussed at the end of this 
section. Defining $h_{2}(d)$ by:
\begin{equation}
h_{2}(d)=\frac{1}{2^{d-2}}\frac{1}{\pi^{\frac{d-1}{2}}}
\frac{1}{\Gamma(\frac{d-1}{2})},
\label{h2}
\end{equation}
it is possible to write $f_{2}(L,m,d,z)|_{m=0}$ as 
\begin{eqnarray}
f_{2}(L,m,d,z)|_{m=0} &=&\frac{1}{2}h_{2}(d)\int_{0}^{\infty}
dk\, k^{d-3}\coth kL \cosh 2kz\nonumber \\
&-&h_{2}(d)\int_{0}^{\infty}
dk\, k^{d-3}\cosh kz\sinh kz.
\label{ult1}
\end{eqnarray}
In a general way, the regularization process is achieved 
introducing exponential 
cut-off regulators and after this the identification of the poles 
of the regularized quantity by means of the Laurent series expansion 
around some point i.e. the negative power portion of such series.
Note that instead of 
imposing renormalization conditions over the 1PI correlation 
functions we can simply subtract the singular part of the Laurent series 
around some point, by the introduction of the counterterms. 
Let us assume $z\neq 0$ and $z\neq L$. A straightforward calculation gives
\begin{eqnarray}
f_{2}(L,m,d,z)|_{m=0} &=&\frac{1}{2}h_{2}(d)\left[\int_{0}^{\infty}
dk\, k^{d-3}(\coth kL-1)\cosh 2kz\right.\nonumber \\
&+& \left.\int_{0}^{\infty}
dk\, k^{d-3}(\cosh 2kz-\sinh 2kz)\right].
\label{ult11}
\end{eqnarray}
In the first integral for large $z$, $(\coth kL-1)$ has the behavior: 
$(\coth kL-1) \sim e^{-2kL}$. Moreover,
the second integral in the above equation is ultraviolet finite for 
$z\neq 0$. Let us define $x=kL$ and $q=kz$ in the first and second 
integrals above respectively. Then Eq.(\ref{ult11})  becomes:
\begin{eqnarray}
f_{2}(L,m,d,z)|_{m=0} &=&\frac{1}{2}h_{2}(d)\frac{1}{L^{d-2}}\int_{0}^{\infty}
dx \,x^{d-3}(\coth x -1) \cosh(\frac{2z}{L}x)\nonumber \\
&+& \frac1{2}h_{2}(d)\frac{1}{z^{d-2}}\int_{0}^{\infty}
dq \,q^{d-3}(\cosh 2q-\sinh 2q).
\label{ult2}
\end{eqnarray}
The second term in the above equation gives us the well known result that 
for a massless minimal coupled scalar field $<\varphi^{2}(x)>$ 
diverges as $\frac{1}{z^{2}}$ if we approach the plate \cite{21}. 
In order to analyze the polar part of $f_{2}(L,0,d,z)$,
we use the definition of the Gamma function. Let us define
$I_{1}(\nu,\mu)$ and $I_{2}(\mu,\beta)$ by
\begin{equation}
I_{1}(\mu,\nu)=\int_{0}^{\infty}
dx \,x^{\mu-1}e^{-\nu x}=\frac1{\nu^{\mu}}\Gamma(\mu),
\,\,\,\,Re(\mu)>0,\,\,\,Re(\nu)>0
\label{I1}
\end{equation}
and 
\begin{equation}
I_{2}(\mu,\beta)=\int_{0}^{\infty}
dx \,x^{\mu-1}e^{-\beta x}(\coth x-1)=2^{1-\mu}\Gamma(\mu)
\zeta(\mu,\frac{\beta}{2}+1)\,\,\,\,Re(\beta)>0,\,\,\,Re(\mu)>1,
\label{I2}
\end{equation}
where $\zeta(z,a)$ is the Riemman zeta function defined by \cite{grads}
\begin{equation}
\zeta(z,a)=\sum_{n=0}^{\infty}\frac{1}{(n+a)^{z}},\,\,\,\,Re(z)>1,
\,\,\,\,\, a \neq 0,-1,-2...
\label{na}
\end{equation}
Then, using Eqs. (\ref{I1}), (\ref{I2}) and (\ref{na}) in Eq. (\ref{ult2})
we have that:
\begin{eqnarray}
f_{2}(L,m,d,z)|_{m=0}&=&\frac{1}{2}h_{2}(d)\frac{1}{L^{d-2}}
\left[2^{2-d}\Gamma(d-2)\left(\zeta(d-2,\frac{z}{L}+1)+
\zeta(d-2,-\frac{z}{L}+1)\right)\right]\nonumber \\
&+& \frac{1}{(2z)^{d-2}}h_{2}(d)\Gamma(d-2).
\label{fim}
\end{eqnarray}
Using the definition of the zeta function, it is evident that:
\begin{eqnarray}
& &\frac{1}{L^{d-2}}
\left(\zeta(d-2,\frac{z}{L}+1)+
\zeta(d-2,-\frac{z}{L}+1)\right)=\nonumber \\
& &\frac{1}{L^{d-2}}\sum_{n=0}^{\infty}
\frac{1}{\left(n+(1+\frac{z}{L})\right)^{d-2}}+\frac1{(L-z)^{d-2}}+
\frac{1}{L^{d-2}}\sum_{n=1}^{\infty}
\frac{1}{\left(n+(1-\frac{z}{L})\right)^{d-2}}.
\label{fim2}
\end{eqnarray}
We see that the regularized
$f_{2}(L,0,d,z)$ has two poles of order $(d-2)$ in $z=0$ and in $z=L$.
Note that the residues of the poles in $z=0$ and in $z=L$ are  
$L$-independent. Since the domain of analyticity of the zeta function is
$d>3$, the case $d=3$ must be studied separately. 
Different treatments for $d=3$ 
and $d=4$ simply express the fact that infrared divergences are more
severe in lower dimensions.

We will go back to Eq.(\ref{f2}), studying the
case $m^2\neq 0$, to see how the IR divergences pop up in the 
$m^2\rightarrow 0$ limit. It is important to stress that, only in the N-N B.C.'s case
we have IR divergences for massless fields, coming from the term $n=0$, i.e., equations (\ref{f1f2}) and (\ref{f2}) are IR finite for $m=0$. 
A straightforward calculation yields
\begin{equation}
f_{2}(L,m,d,z)=\frac{1}{2}h_2(d)
\int_{0}^{\infty} d\rho\frac{\rho^{d-2}}
{({\rho}^{2}+m^{2})^{\frac{1}{2}}}
\frac{\cosh((L-2z)({\rho}^{2}+m^{2})^{\frac{1}{2}})}
{\sinh (L({\rho}^{2}+m^{2})^{\frac{1}{2}})}.
\label{f21}
\end{equation}
Defining $\sigma=({\rho}^{2}+m^{2})^{\frac{1}{2}}$ and using the fact
that $d=3$, we have:
\begin{eqnarray}
f_{2}(L,m,d,z)|_{d=3} &=&\frac{1}{2}h_{2}(3)\left[\int_{m}^{\infty}
d\sigma\, (\coth \sigma L-1)\cosh 2\sigma z\right.\nonumber \\
&+& \left.\int_{m}^{\infty}
d\sigma\, (\cosh 2\sigma z-\sinh 2\sigma z)\right].
\label{f22}
\end{eqnarray}
The second integral in the above expression is convergent
for $z\neq 0$, and defining $v=2\sigma z$, it becomes:
\begin{equation}
\frac1{4z}h_{2}(3)\int_{2mz}^{\infty} dv\,e^{-v}=
 \frac1{4z}h_{2}(3)\Gamma(1,2m z),
\label{f23}
\end{equation}
where $\Gamma(a,x)$ is the incomplete gamma function. Consequently,
we have a simple pole for $z=0$. Again, the residue of this pole is
$L$-independent. To complete the regularization procedure we have now
to analyze the first integral of Eq.(\ref{f22}):
\begin{equation}
 \frac{1}{2}h_{2}(3)\int_{m}^{\infty}
 d\sigma\, (\coth \sigma L-1)\cosh 2\sigma z=
 \frac1{4L}h_2(3)\int_{mL}^{\infty} du \frac{e^{\frac{z}{L}u}}{e^u-1}+
\frac1{4L}h_2(3)\int_{mL}^{\infty} du \frac{e^{-\frac{z}{L}u}}{e^u-1}.
\label{f24}
\end{equation}
The second integral in the right side of Eq.(\ref{f24}) is convergent
and the first one has a simple pole at $z=L$, again with an $L$-independent
residue. 

From the discussion above, we can conclude that, in order to eliminate 
the ultraviolet divergences of the theory we have to introduce counterterms as surface interactions, and consequently
the full action will have the following form for both fields $\varphi_{1}$ 
and $\varphi_{2}$:
\begin{equation}
S(\varphi)=\int_{0}^{L}dz \int d^{d-1}r
(\frac{1}{2}(\partial\varphi)^{2}+\frac{1}{2}m^2\varphi^2+
\frac{1}{4!}\lambda\varphi^{4})+
\int d^{d-1}r(c_{1}\varphi^{2}(\vec{r},0)+c_{2}\varphi^{2}(\vec{r},L)).
\label{sur}
\end{equation}

As we have already taken care of
the ultraviolet divergences, let us study the infrared divergent 
piece (for $m=0$) of $f_{2}(L,m,d,z)|_{d=3}$. Let us call this piece $f^{*}_{2}(L,m,d,z)|_{d=3}$. Note that we introduce an ultraviolet 
cut-off in order to use the Bernoulli representation of the integrand.
\begin{equation}
f^{*}_{2}(L,m,d,z)|_{d=3}=
\frac1{4L}h_2(3)\left(
\int_{m L}^{2\pi} du \frac{e^{\frac{z}{L}u}}{e^u-1}+
\int_{mL}^{2\pi} du \frac{e^{-\frac{z}{L}u}}{e^u-1}\right).
\label{f25}
\end{equation}
Writing the integrand using the Bernoulli polynomials it is not difficult
to show that 
\begin{equation}
f^{*}_{2}(L,m,3,z)=\frac1{2L}h_2(3)B_0(\frac{z}{L})ln(\frac{2\pi}{mL})
+regular\,\,part\,\,(f^{*}_{2}(L,m,z)).
\label{f26}
\end{equation}
When $m\rightarrow 0$ we have a logarithmic divergence which is $z$ 
dependent.

Going back to the case of the ultraviolet divergence,
some authors claimed that the introduction of surface counterterms is 
against the spirit of the renormalization program. In our case, however, 
it is possible to change the model, by adding a new interaction term, in 
such a way that the ultraviolet divergences coming from the $f_{2}(L,m,d,z)$
contributions corresponding to each field are compensated.
One possibility is to consider the $O(2)$ symmetric model with a 
$\varphi_{1}^{2}\varphi_{2}^{2}$ interaction term.  Then the  
$z$-dependent part of each tadpole cancel each other out. The situation is
similar to the case of supersymmetric theories, where the finiteness of some correlation functions is achieved by a balance between bosonic and fermionic 
loops. 
The $O(2)$ symmetric model with an interaction 
term $\varphi_{1}^{2}\varphi_{2}^{2}$ ($DD$ and $NN$ B.C.'s) developes a 
size dependent mass $\Delta m^{2}$ proportional to $gL^{-2}$, as the 
$\varphi^{4}$ model at finite temperature. In the same way as temperature can solve
the IR problem in some QFT models, finite size effects can also cure these divergences,
when a resummation can be implemented.

As stressed by many authors, in the electromagnetic 
case, the  origin of the unboundedness renormalized stress-tensor near a curved 
surface has the origin in the unphysical nature of classical 
"perfect conductor" boundary condition. Let us suppose the following 
physical situation. For the low energy modes the manifold is 
$[0,L]\times \Re^{d-1}$ and for the high energy modes we have 
$S^{1}\times \Re^{d-1}$, i.e. let us assume a sharp cut-off and for 
$k_{n}< \Lambda$ we have $DD$ boundary conditions and 
for $k_{n}\geq \Lambda$ we have periodic B.C.'s. The high frequencies
do not "see" the mirror at $z=0$, and translational invariance is 
maintained only for these modes. If the collapse of the 
renormalization program (removing infinities from 
perturbative calculations using only "bulk" counterterms) is related 
with the break of translational 
invariance, our improved model must be renormalizable. A further 
study of this model may be of interest. A different possibility is to construct an effective action for the slow-modes and after 
this imposing the $DD$ or $NN$ b.c. \cite{Wilson}\cite{He}.

\section{Conclusions}

In this paper we studied finite size effects in an interacting field 
theory, with broken  translation invariance.
We calculated the vacuum activity for an anisotropic model,
between two parallel plates in a $d$ dimensional Euclidean space. It has been
possible to obtain closed expressions for $<\varphi^{2}_{1}(x)>$ and $<\varphi^{2}_{2}(x)>$, for fields satisfying $DD$ and $NN$ B.C.'s, respectively.  
For different shapes, the complexity and the number of technical difficulties
increase enormously, but the multiple scattering method can be 
used in these cases \cite{balian}, at least for small curvatures of the 
boundaries. 
We presented a model having  the interesting property that 
the $z$-dependent part of the tadpole graphs for $DD$ and $NN$ B.C.'s 
have the same modulus and opposite signs. 
This fact could explain the boundedness of the renormalized vacuum 
expectation value of the energy-stress tensor of the electromagnetic field 
in the Casimir-like configuration. 

We have also seen that, to renormalize the theory,  counterterms corresponding to surface interactions are required. 
One can, however,  avoid this difficulty by equipping the model 
with a $\varphi_{1}^{2}\varphi_{2}^{2}$ interaction. 
Then, the $z$ dependent pieces of each tadpole cancel each other out, and the two fields develop a size dependent mass $\Delta m^{2}$ proportional to $gL^{-2}$, as for the  single $\varphi^{4}$ model at finite temperature.

There are several directions in which the finite size effects for systems 
with breaking of translational invariance which may deserve further research. 
To mention a few of them: the study of interacting fermions, 
the non-linear $\sigma$ model in domains with 
one finite direction and $(d-1)$ infinite directions~\cite{floratos}, 
and finally as a straightforward extension of this 
work, the study of the $O(2)$ symmetric model at the two-loop approximation. 

As discussed in the Introduction, one should  prove that the 
the renormalization program can be implemented beyond the one-loop approximation, where overlapping divergences emerge.

\section{Acknowlegements}

We would like to thank B.Schroer and R.De Paola for several helpful 
discussions. We are also grateful to B.F.Svaiter for very useful 
discussions and comments. N.F.Svaiter would like to acknowledge the
hospitality of the Centro At\'omico Bariloche where part of this
work was carried out.This paper was supported by Conselho Nacional de 
Desenvolvimento Cientifico e Tecnologico do Brazil (CNPq) and 
Centro Latino Americano de Fisica (CLAF).

\end{document}